\documentclass{article}
\usepackage{amssymb}

\newcommand{\Ket}[1]{\left\vert#1\right\rangle}

\newcommand{\KetBra}[2]{\left\vert#1\right\rangle\left\langle#2\right\vert}
\newcommand{\ho}{\hat{H}_\mathrm{u}}
\newcommand{\hi}{\hat{H}_\mathrm{p}}
\newcommand{\hl}{\hat{H}(\lambda)}
\newcommand{\hlt}{\hat{K}(\lambda)}
\newcommand{\rotham}{\hat{\mathfrak{H}}}
\newcommand{\HS}{\hat{H}_{\updownarrow}}
\newcommand{\app}{\stackrel{\ \lambda^2}{\approx}}
\newcommand{\evo}{\hat{T}_\Lambda}
\newcommand{\BsId}{1_B}

\begin{document}

\title{Coarse grained and fine dynamics in trapped ion Raman schemes}
\author{
B. Militello$^{1}$\footnote{bdmilite@fisica.unipa.it}, P.
Aniello$^{2}$\footnote{Paolo.Aniello@na.infn.it}, and A.
Messina$^{1}$\footnote{messina@fisica.unipa.it}
\\
$^{1}$ {\small Istituto Nazionale di Fisica della Materia,
Unit\`{a} di Palermo, MIUR}\\{\small and Dipartimento di Scienze
Fisiche ed Astronomiche dell'Universit\`{a} di Palermo,}
\\ {\small Via Archirafi, 36 - 90123 Palermo, Italy}
\\
$^{2}$ {\small Istituto Nazionale di Fisica Nucleare, Sezione di
Napoli,}\\{\small and Dipartimento di Scienze Fisiche
dell'Universit\`{a} di Napoli \lq\lq Federico II\rq\rq},\\
{\small Complesso Universitario di Monte S. Angelo, Via Cintia -
80126 Napoli, Italy}}


\maketitle

\begin{abstract}

A novel result concerning Raman coupling schemes implemented using
trapped ions is obtained. By means of an operator perturbative
approach, it is shown that the complete time evolution of these
systems can be expressed, with a high degree of accuracy, as the
product of two unitary evolutions. The first one describes the
time evolution related to an effective coarse grained dynamics.
The second is a suitable correction restoring the {\em fine}
dynamics suppressed by the coarse graining performed to
adiabatically eliminate the nonresonantly coupled atomic level.
\end{abstract}

\vspace{2 cm}


\clearpage

\section{Introduction}

Trapped ions provide an effective platform for observing
interesting aspects of quantum mechanics and for realizing useful
applications in the context of quantum
computation~\cite{Wine_Blatt,nist98,nonclassical-science}.

In these traps, a time-dependent quadrupolar electromagnetic
field is responsible for a charged particle motion which may be
kinematically assimilated to the motion of a massive spot
subjected to a quadratic potential. Such a circumstance provides
the possibility of describing the center of mass of an ion
confined into a rf Paul trap as a quantum harmonic
oscillator~\cite{Toschek,Ghosh,Knight-review}. In addition the ion
possesses atomic degrees of freedom related to the electronic
motion around the
nucleus~\cite{nist98,Toschek,Ghosh,Knight-review,Vogel-review}.

Acting upon the system via laser fields, it is possible to induce
vibronic transitions described by Jaynes-Cummings-like
Hamiltonians~\cite{Vogel-review,non-lin-JC}. Such interactions are
characterized by nonlinearities governed by the so called
Lamb-Dicke parameter which, in a spherically symmetric trap,
is nothing but the ratio between the
width of the ion vibrational ground wave function and the laser
wavelength. Controlling the Lamb-Dicke parameter leaving unchanged
the laser frequency would then provide the possibility of
implementing a wider variety of Hamiltonian models. Unfortunately, these
two parameters are strictly related, the Lamb-Dicke parameter
being proportional to the inverse of the wavelenght, hence
proportional to the laser frequency.
Therefore, they cannot be independently adjusted. Nevertheless,
such a possibility can be achieved exploiting a fundamental property of
Raman coupling schemes. In these
couplings a three-level system is subjected to a far detuned
$\Lambda$-scheme. Under some assumptions and within approximations
concerning the time scale used to observe the system, the dynamics
may be well enough described via an effective Hamiltonian
thinkable as a Jaynes-Cummings-like Hamiltonian
related to an {\em effective}
laser field, with frequency and wave-vector given by the differences
between the corresponding parameters of the two {\em real}
$\Lambda$-scheme lasers. Then, changing the angle between the two
real laser propagation directions leads to the possibility of
obtaining effective lasers such that the product of
their wavelenght and frequency is not the velocity of light.
The price to pay is the complete ignorance
about the detailed system dynamics at a {\em fine} time scale.

In this paper we will try to overcome such a limit by means of a
new approach. More precisely, we will analyze the dynamics of a
three-level trapped ion subjected to a Raman $\Lambda$ coupling
scheme using a very convenient perturbative decomposition of the
evolution operator of the system. We will show that the Raman
scheme time evolution can be factorized, at the second
perturbative order (hence with a high degree of accuracy), into
two unitary evolutions. The first one can be interpreted as the
effective time evolution which may be obtained adiabatically
eliminating one of the three atomic levels involved in the
coupling scheme~\cite{Steinbach}. Such a dynamics concerns the
{\em coarse grained} variables and is in accordance with already
well known results~\cite{Steinbach}. The second unitary evolution
introduces the correction necessary to take into account the {\em
fine} deviation from the coarse grained time evolution.

\section{The physical system}

The physical system on which we focus is a three-level harmonically
trapped ion
subjected to a Raman coupling scheme involving atomic transitions.
The relevant Schr\"odinger picture Hamiltonian of the Raman
$\Lambda$-scheme is then given by
\begin{eqnarray}
\label{Raman_Hamiltonian_SchrPct} \nonumber
\!\!\!\hat{H}_{\Lambda}(t)
\!\!\!&=&\!\!\!\sum_{l=1,2,3}\hbar\omega_l\hat{\sigma}_{ll}+
\hbar\nu\sum_{\alpha=x,y,z}\hat{a}_{\alpha}^{\dag}
\hat{a}_{\alpha}^{\phantom{\dagger}}\\\!\!\!&+&\!\!\!
\left[\hbar
g_{13}e^{-i\left(\vec{k}_{13}\cdot\vec{r}-\omega_{13}t\right)}\hat{\sigma}_{13}+h.c.\right]+
\left[\hbar
g_{23}e^{-i\left(\vec{k}_{23}\cdot\vec{r}-\omega_{23}t\right)}\hat{\sigma}_{23}+h.c.\right],
\end{eqnarray}
where $\hat{\sigma}_{kl}\equiv\KetBra{k}{l}$ (with $k,l=1,2,3$),
$\{\Ket{l}\}$ being the considered three atomic levels and
$\{\hbar\omega_l\}$ the corresponding energies;
$\hat{a}_{\alpha}^{\phantom{\dagger}}$
($\alpha=x,y,z$) is the annihilation operator related to the
center of mass harmonic motion along the direction $\alpha$
(we will denote the associated Fock basis by $\{\psi_n^\alpha\}$).
For the sake of
simplicity (but without loss of generality), we have assumed to
deal with a 3D degenerate parabolic trap with single frequency $\nu$.
The two laser fields responsible for
the coupling terms are characterized by complex strengths (proportional to
the laser ampliutude and to the atomic dipole operator, and
including the laser phases),
wave vectors and frequencies $g_{13}$, $\vec{k}_{13}$,
$\omega_{13}$ and $g_{23}$, $\vec{k}_{23}$, $\omega_{23}$
respectively.

The level $\Ket{3}$ is assumed to be dipole-coupled to both the levels
$\Ket{1}$ and $\Ket{2}$ via far detuned lasers. Precisely, the two
laser frequencies are chosen in such a way that
\begin{eqnarray}\label{Detuning}
  \Delta\equiv\omega_3-\omega_1-\omega_{13}=\omega_3-\omega_2-\omega_{23},
\end{eqnarray}
where the detuning $\Delta$ satisfies the condition
\begin{eqnarray}\label{Detuning_Conditions}
  |\Delta|>>|g_{13}|,|g_{23}|,\nu.
\end{eqnarray}

The analysis of such an Hamiltonian model has been already carried
out, for instance in ref.~\cite{Steinbach}, providing the
adiabatic elimination of the nonresonantly coupled atomic level
$\Ket{3}$ following the path pointed out in
refs.~\cite{Allen_Stroud,Shore} and including the motional degrees
of freedom. Indeed, due to the large detuning, the transitions for
instance from the level $\Ket{1}$ to the level $\Ket{3}$ are very fast and
immediately followed by decays on the atomic level $\Ket{2}$.
Therefore, considering only coarse grained observables, meaning that the
system is observed at a \lq\lq rough enough time scale\rq\rq,
effectively eliminates the far detuned level; namely, at
such a time scale, the only observables and hence meaningful
dynamical behaviors, involve levels $\Ket{1}$ and $\Ket{2}$ as a
result of time averaging second order processes having $\Ket{3}$
as an intermediate virtual level. This procedure hence suppresses
the {\em fine} dynamics, that is it sacrifices any information
concerning the fast dynamics the third level is involved in.

In the following we present a perturbative approach to the
solution of the dynamical problem related to $\hat{H}_{\Lambda}$
overcoming the limit of the coarse graining, making it possible to
study also the {\em fine} dynamics discarded by the adiabatic
elimination. The first step consists in passing to a rotating
frame, meaning that the time-dependent Hamiltonian
$\hat{H}_\Lambda$ is canonically transformed via the operator
\begin{equation}\label{Rotating_Transformation}
\hat{R}(t)=e^{-i\hat{A}t},
\end{equation}
where
\begin{eqnarray}
\hat{A}\!\!\!& =& \!\!\! \left(\omega_3-\Delta\right)
\left(\hat{\sigma}_{11}+\hat{\sigma}_{22}+\hat{\sigma}_{33}\right)
-\omega_{13}\hat{\sigma}_{11}-\omega_{23}\hat{\sigma}_{22}\nonumber\\
\!\!\! & = & \!\!\! \label{Rotating_Transformation_Generator}
\omega_1\hat{\sigma}_{11}+\omega_2\hat{\sigma}_{22}+\left(\omega_3-\Delta\right)\hat{\sigma}_{33},
\end{eqnarray}
into the following time-independent rotating frame Hamiltonian:
\begin{equation}\label{Raman_Hamiltonian}
\rotham := \hat{R}(t)^\dagger
\left(\hat{H}_\Lambda(t)-\hat{A}\right) \hat{R}(t)= \hbar\Delta\,
\hat{H} = \hbar\Delta\left(\hat{H}_0+\hat{H}_B+\HS\right),
\end{equation}
where $\hat{H}$ is a {\it dimensionless Hamiltonian} which is the
sum of the three hermitian operators $\hat{H}_0$, $\hat{H}_B$ and
$\HS$ defined as
\begin{equation}\label{Raman_Hamiltonian_Explenation}\nonumber
\cases{ \hat{H}_0\,:=&$\!\!\hat{\sigma}_{33},$\cr
\hat{H}_B:=&$\!\!\frac{\nu}{\Delta}\sum_{\alpha=x,y,z}\hat{a}_{\alpha}^{\dag}\hat{a}_{\alpha}$,\cr
\HS\,\,
:=&$\!\!\left[\frac{g_{13}}{\Delta}e^{-i\vec{k}_{13}\cdot\vec{r}}\hat{\sigma}_{13}+h.c.\right]+\left[\frac{g_{23}}{\Delta}
e^{-i\vec{k}_{23}\cdot\vec{r}}\hat{\sigma}_{23}+h.c.\right]$.\cr }
\end{equation}

Considering the assumption given by the
inequality~(\ref{Detuning_Conditions}), both $\hat{H}_B$ and $\HS$
may be thought of as perturbations with respect to $\hat{H}_0$. In
fact, introducing the dimensionless perturbative parameter
\begin{equation}
\lambda:= \frac{g}{\Delta},\ \ \ g\equiv\max\{\nu,
|g_{13}|,|g_{23}|\},
\end{equation}
both $\hat{H}_B$ and $\HS$ are first order perturbations in
$\lambda$:
\begin{equation}
\hat{H}=\hl=\hat{H}_0
+\lambda\,\varkappa\sum_{\alpha=x,y,z}
\hat{a}_{\alpha}^{\dag}\hat{a}_{\alpha}^{\phantom{\dagger}}
+\lambda \sum_{j=1,2}
\left[\varkappa_{j3}\,e^{-i\vec{k}_{j3}\cdot\vec{r}}\,\hat{\sigma}_{j3}+h.c.\right],
\end{equation}
where $\varkappa\equiv\nu/ g\le 1$, $\varkappa_{j,3}\equiv
g_{j,3}/ g$, $|\varkappa_{j,3}|\le 1$, and we notice that, due to
condition~(\ref{Detuning_Conditions}), $\lambda\ll 1$. It is worth
noting that the circumstance that $\hat{H}_B$ is treated as a
perturbation leads to the {\em eccentric} situation of an
unperturbed Hamiltonian, $\hat{H}_0$, wherein the bosonic degrees
of freedom are {\em absent}. Nevertheless, as we shall see, such a
mathematical artifice reveals fruitful in order to succeed in
factorizing the coarse grained dynamics and its fine correction.

Our solving procedure relies on a suitable canonical
transformation $e^{i\hat{Z}(\lambda)}$ of the rotating frame
Hamiltonian such that
\begin{equation}\label{diagonalization}
e^{i\hat{Z}(\lambda)}\,\rotham\, e^{-i\hat{Z}(\lambda)}=\hbar\Delta\,
e^{i\hat{Z}(\lambda)}\,\hat{H}(\lambda)\,e^{-i\hat{Z}(\lambda)}=
\hbar\Delta\left(\hat{H}_0+\hat{C}(\lambda)\right),
\end{equation}
where $\hat{C}(\lambda), \hat{Z}(\lambda)$ depend analytically on
the perturbative parameter $\lambda$ and $\hat{C}(\lambda)$ is a
{\it constant of motion} with respect to the unperturbed dynamics,
i.e. $[\hat{H}_0,\hat{C}(\lambda)]=0$. This transformation allows
to give a very convenient decomposition of the evolution operator
associated with the rotating frame Hamiltonian, namely
\begin{equation} \label{evol}
\exp\!\left(-\frac{i}{\hbar}\rotham\,t\right)=
e^{-i\hat{Z}(\lambda)}\,\exp(i\Delta\,\hat{H}_0\,
t)\,\exp(i\Delta\,\hat{C}(\lambda)\,t)\,e^{i\hat{Z}(\lambda)}.
\end{equation}
At this point, truncating the power expansions
\[
\hat{C}(\lambda)=\lambda\,\hat{C}_1+\lambda^2\hat{C}_2+\cdots+\lambda^n\hat{C}_n+\cdots,\
\
\hat{Z}(\lambda)=\lambda\,\hat{Z}_1+\lambda^2\hat{Z}_2+\cdots+\lambda^n\hat{Z}_n+\cdots
\]
at a given perturbative order, one obtains by
formula~{(\ref{evol})} useful expressions of the evolution
operator. This procedure has been developed in a general setting
in refs.~\cite{Aniello1,Aniello2}. In the next section, we want to
recall briefly the mathematical background and to show how the
operators $\hat{C}_1,\hat{C}_2,\ldots$,
$\hat{Z}_1,\hat{Z}_2,\ldots$ can be computed by a suitable
iterative process. Then, we will give the explicit solutions for
our case up to the second perturbative order.

\section{Perturbative analysis of the rotating frame Hamiltonian}

Let $\ho$, $\hi$ be hermitian operators and assume that $\ho$ has
a purely discrete spectrum. Denote by $ E_0 < E_1 < E_2 < \ldots $
the (possibly degenerate) eigenvalues of $\ho$ and by $\hat{P}_0, \hat{P}_1,
\hat{P}_2,\ldots$ the associated eigenprojectors. Now, consider the
operator $ \hl=\ho+\lambda\,\hi, \ \lambda\in\mathbb{C}, $ which
is hermitian if $\lambda$ is real. It is possible to show that,
under certain conditions~\cite{Kato}, there exist positive
constants $r_0,r_1,r_2,\ldots$ and a simply connected
neighbourhood $\mathcal{I}$ of zero in $\mathbb{C}$ such that the
following contour integral on the complex plane
\[
\hat{P}_m(\lambda)=\frac{i}{2\pi}\oint_{|E-E_m|=r_m}\!\!\!\!\!\!\!\!\!\!\!\!\!
\!\!\!\!\!\!\!\!\!\! dE\ \ \ \ \ \left(\hl-E\right)^{-1}  \ \ \ \
\lambda\in\mathcal{I},
\]
defines a projection ($\hat{P}_m(\lambda)^2=\hat{P}_m(\lambda)$), which is an
orthogonal projection for real $\lambda$, with $\hat{P}_m(0)=\hat{P}_m$, and
$\mathcal{I}\ni\lambda\mapsto \hat{P}_m(\lambda)$ is an analytic
operator-valued function. Moreover, the range of $\hat{P}_m(\lambda)$ is
an invariant subspace for $\hl$, hence
\begin{equation} \label{inv}
\hl\, \hat{P}_m(\lambda) = \hat{P}_m(\lambda)\, \hl\, \hat{P}_m(\lambda),
\end{equation}
and there exists an analytic family $\hat{U}(\lambda)$ of invertible
operators such that
\begin{equation} \label{trasf}
\hat{P}_m = \hat{U}(\lambda)\, \hat{P}_m(\lambda) \, \hat{U}(\lambda)^{-1},\ \ \
\hat{U}(0)=\mathrm{Id},
\end{equation}
and $ \hat{U}(\lambda)=e^{i\hat{Z}(\lambda)},\
\lambda\in\mathcal{I}, $ with
$\hat{Z}(\lambda^*)=\hat{Z}(\lambda)^\dagger$ (hence, for real $\lambda$,
$\hat{Z}(\lambda)$ is hermitian and $\hat{U}(\lambda)$ is unitary), where
$\mathcal{I}\ni\lambda\mapsto \hat{Z}(\lambda)$ is analytic. One can
show easily that the function $\lambda\mapsto \hat{U}(\lambda)$ is not
defined uniquely by condition~{(\ref{trasf})} even in the simplest
case when $\ho$ has a nondegenerate spectrum. Anyway, the nonunicity in the
definition of $\hat{U}(\lambda)$ is not relevant if one is only
interested in obtaining an expression of the evolution operator associated
with $\hl$ of the general form~{(\ref{evol})}. We will see soon that there
is a natural condition which fixes a unique solution for $\hat{U}(\lambda)$.

Now, let us define
the operator
\begin{equation}
\hlt:=\hat{U}(\lambda)\, \hl\, \hat{U}(\lambda)^{-1},
\end{equation}
which, for real $\lambda$, is unitarily equivalent to $\hl$. Using
relations~{(\ref{inv})} and (\ref{trasf}), we find
\[
\hlt\, \hat{P}_m = \hat{U}(\lambda)\, \hl\, \hat{P}_m(\lambda)\,
\hat{U}(\lambda)^{-1} =
\hat{U}(\lambda)\, \hat{P}_m(\lambda)\,\hl\,
\hat{P}_m(\lambda)\, \hat{U}(\lambda)^{-1}
\]
and hence: $ \hlt\, \hat{P}_m= \hat{P}_m \,\hlt\, \hat{P}_m\ .$
It follows that $
\left[\ho,\,\hlt\right]=0 $ and then we obtain the following
important decomposition formula:
\begin{equation} \label{decomposition}
\hat{U}(\lambda)\, \hl\, \hat{U}(\lambda)^{-1} = \ho + \hat{C}(\lambda),
\end{equation}
where $\left[\hat{C}(\lambda),\ho\right]=0$, i.e.\
$\hat{C}(\lambda)$ is a constant of the motion with respect to the
time evolution generated by $\ho$. At this point, we can obtain
perturbative expressions of the unknown operators
$\hat{C}(\lambda)$, $\hat{U}(\lambda)$ by means of a recursive
algebraic procedure.\\
Indeed, since the functions $\lambda\mapsto
\hat{C}(\lambda)$ and $\lambda\mapsto \hat{Z}(\lambda)$ are
analytic in $\mathcal{I}$ and $\hat{C}(0)=\hat{Z}(0)=0$, we can
write:
\begin{equation} \label{power}
\hat{C}(\lambda)=\sum_{n=1}^{\infty}\lambda^n\, \hat{C}_n,\ \ \
\hat{Z}(\lambda)=\sum_{n=1}^{\infty}\lambda^n\, \hat{Z}_n \ \ \ \
\lambda\in\mathcal{I}.
\end{equation}
In order to determine the operators $\{\hat{C}_n\}$ and
$\{\hat{Z}_n\}$, we substitute the exponential form
$e^{i\hat{Z}(\lambda)}$ of $\hat{U}(\lambda)$ in
formula~{(\ref{decomposition})} thus getting
\[
\hl + \sum_{n=1}^{\infty} \frac{i^n}{n!}\,
\mathrm{ad}_{\hat{Z}(\lambda)}^n\, \hl= \ho+
\hat{C}(\lambda).
\]
where we recall that
$\mathrm{ad}_{\hat{Z}(\lambda)}\hl:=[\hat{Z}(\lambda),\hl]$.\\
Next, inserting the power expansions~{(\ref{power})} in this
equation, in correspondence to the various perturbative orders, we
obtain the following set of conditions:
\begin{eqnarray*}
\hat{C}_1 -i\left[\hat{Z}_1,\ho\right]-\hi=0 & , & \left[\hat{C}_1,\ho\right]=0
\label{first}
\\
\hat{C}_2-i\left[\hat{Z}_2,\ho\right]
+\frac{1}{2}\left[\hat{Z}_1,\left[\hat{Z}_1,\ho\right]\right]
-i\left[\hat{Z}_1,\hi\right]=0 & , & \left[\hat{C}_2,\ho\right]=0
\label{second}
\\
& \vdots & \nonumber
\end{eqnarray*}
where we have taken into account also the additional constraint
$[\hat{C}(\lambda),\ho]=0$. This infinite set of equations can be
solved recursively. The first equation, together with the first
constraint, determines $\hat{Z}_1$ up to an operator commuting
with $\ho$ and $\hat{C}_1$ uniquely and so on. It is convenient to
eliminate the arbitrariness in the determination of the operators
$\{\hat{Z}_n\}$ choosing the {\em minimal solution} characterized
by the additional condition $\sum_m
\hat{P}_m\hat{Z}_n\hat{P}_m=0,\ n=1,2,\ldots\ $.

In our particular case, we have the following identifications:
\begin{equation}\label{Identification}
  \cases{
    \ho\equiv\hat{H}_0,\cr
    \hi\equiv\hat{H}_B+\HS.
  }
\end{equation}
Notice that the two (infinitely degenerate) eigenspaces of the
unperturbed Hamiltonian $\hat{H}_0$ are associated with the
eigenprojectors
\begin{equation}\label{Deg_Eigenspaces}
  \{\hat{P}_m\}_{m=g,e}=
  \{
  \hat{P}_g\equiv
\BsId\otimes\left(\hat{\sigma}_{11}+\hat{\sigma}_{22}\right)
  \; ,\;
  \hat{P}_e\equiv
\BsId\otimes\left(\hat{\sigma}_{33}\right)
  \},
\end{equation}
where
\begin{equation}\label{Bos_Identity}
 \BsId\equiv\sum_{n_{x},n_{y},n_{z}}
\left(\KetBra{\psi_{n_x}^x}{\psi_{n_x}^x}\right)
 \otimes\cdots
\otimes\left(\KetBra{\psi_{n_z}^z}{\psi_{n_z}^z}\right)
\end{equation}
is the identity in the vibrational Hilbert space.
Accordingly, for the operators
$\{\hat{C}_1,\hat{Z}_1,\hat{C}_2,\hat{Z}_2,\ldots\}$ forming the
minimal solution, we get at the first perturbative order the
following expressions:
\begin{equation}\label{First_Order}
\cases{
\lambda\,\hat{C}_1=&$\!\!\sum_{m=e,g}\hat{P}_{m}\left(\hat{H}_B+\HS\right)\hat{P}_{m}$,\cr
\lambda\,\hat{Z}_1=&$\!\!i\sum_{j\not=k}\hat{P}_{j}\left(\hat{H}_B+\HS\right)\hat{P}_{k}$.\cr
}
\end{equation}
Similarly, at the second order, we have:
\begin{equation}\label{Second_Order}
\cases{ \lambda^2\hat{C}_2=&$\!\!\sum_{m=e,g}\hat{P}_{m}\left\{
                                   i\left[\hat{Z}_1, \hat{H}_B+\HS\right]-\frac{1}{2}\left[\hat{Z}_1,[\hat{Z}_1,\hat{H}_0]\right]
                                   \right\}\hat{P}_{m}$,\cr
\lambda^2\hat{Z}_2=&$\!\!i\sum_{j\not=k}\hat{P}_{j}\left\{
                                   i\left[\hat{Z}_1, \hat{H}_B+\HS\right]-\frac{1}{2}\left[\hat{Z}_1,[\hat{Z}_1,\hat{H}_0]\right]
                                   \right\}\hat{P}_{k}$.\cr
}
\end{equation}
Eventually, performing explicit calculations, we find that

\begin{equation}\label{C1}
  \lambda\hat{C}_1=\hat{H}_B,
\end{equation}

\begin{eqnarray}
  \lambda^2\hat{C}_2 \!\!\! & = & \!\!\! -\frac{|g_{13}|^2}{\Delta^2}\hat{\sigma}_{11}
            -\frac{|g_{23}|^2}{\Delta^2}\hat{\sigma}_{22}
            +\frac{|g_{13}|^2+|g_{23}|^2}{\Delta^2}\hat{\sigma}_{33}
\nonumber \\ \label{C2}
\!\!\!   & - & \!\!\! \left(
               \frac{g_{13}g_{32}}{\Delta^2}e^{-i\vec{k}_{13}\cdot\vec{r}}e^{i\vec{k}_{23}\cdot\vec{r}}\hat{\sigma}_{12}
               +h.c.
             \right),
\end{eqnarray}
where we have set $g_{3j}\equiv g_{j3}^*$, and
\begin{equation}\label{Z1}
  \lambda\hat{Z}_1=i\left(
\frac{g_{13}}{\Delta}e^{-i\vec{k}_{13}\cdot\vec{r}}\hat{\sigma}_{13}-h.c.\right)+i\left(\frac{g_{23}}{\Delta}e^{-i\vec{k}_{23}\cdot\vec{r}}\hat{\sigma}_{23}-h.c.
             \right),
\end{equation}

\begin{equation}\label{Z2}
  \lambda^2\hat{Z}_2=\frac{\nu}{\Delta}
            \left\{
\left(\frac{g_{13}}{\Delta}
\hat{X}_{13}\hat{\sigma}_{13}+\frac{g_{31}}{\Delta}
\hat{X}_{31}\hat{\sigma}_{31}\right)+
\left(\frac{g_{23}}{\Delta}
\hat{X}_{23}\hat{\sigma}_{23}+\frac{g_{32}}{\Delta}
\hat{X}_{32}\hat{\sigma}_{32}\right)
            \right\},
\end{equation}
where:

\begin{eqnarray}\label{A13}\nonumber
\cases{
 \hat{X}_{j3}:=&
  $\!\!i[e^{-i\vec{k}_{j3}\cdot\vec{r}},\sum_{\alpha=x,y,z}\hat{a}_{\alpha}^{\dag}\hat{a}_{\alpha}^{\phantom{\dagger}}]$,\cr
\hat{X}_{3j}:=&
$\!\!i[e^{i\vec{k}_{j3}\cdot\vec{r}},\sum_{\alpha=x,y,z}\hat{a}_{\alpha}^{\dag}\hat{a}_{\alpha}^{\phantom{\dagger}}]=\hat{X}_{j3}^\dagger$,\cr
}
\end{eqnarray}
with $j=1,2$.

The interpretation of this result leads to a very interesting
fact. Indeed, it turns out that once the unitary transformation
$e^{iZ(\lambda)}$ has been applied to the rotating frame
Hamiltonian $\rotham$ (recall eq.~{(\ref{diagonalization})}), the
time evolution of the system is described, at the second order in
the parameter $\lambda$, by the Hamiltonian
\begin{equation}\label{Dressed_Hamiltonian}
\hbar\Delta\left(\hat{H}_0+\lambda\hat{C}_1+\lambda^2\hat{C}_2\right)
= \rotham_{12} +\rotham_{3},
\end{equation}
where, in order to display a more transparent formula, we set
\begin{eqnarray}
\rotham_{12}
\!\!\! &:=& \!\!\!
\hbar\nu\sum_{\alpha=x,y,z}(\hat{a}_\alpha^\dagger
\hat{a}_\alpha^{\phantom{\dagger}})
\otimes(\hat{\sigma}_{11}+\hat{\sigma}_{22})+
\hbar\breve{\omega}_1\hat{\sigma}_{11}
+\hbar\breve{\omega}_2\hat{\sigma}_{22}
\nonumber \\
\!\!\! & + & \!\!\!
\left[\hbar g_{12}e^{-i \vec{k}_{12}\cdot\vec{r}}\hat{\sigma}_{12}
+ h.c.\right],
\end{eqnarray}
\begin{equation}
\rotham_{3}\, :=\,
\hbar\nu\sum_{\alpha=x,y,z}(\hat{a}_\alpha^\dagger
\hat{a}_\alpha^{\phantom{\dagger}})\otimes\hat{\sigma}_{33}+
\hbar(\Delta+\breve{\omega}_3)\hat{\sigma}_{33},
\end{equation}
with:
\[
\breve{\omega}_{j}=-\frac{|g_{j3}|^2}{\Delta},\ j=1,2\, ,\
\breve{\omega}_3=\frac{|g_{13}|^2+|g_{23}|^2}{\Delta},\
g_{12}=\frac{g_{13}g_{32}}{\Delta},\
\vec{k}_{12}=\vec{k}_{13}-\vec{k}_{23}.
\]
Thus, the transformed Hamiltonian is the sum of two decoupled
Hamiltonians $\rotham_{12}$ and $\rotham_3$,
$[\rotham_{12},\rotham_3]=0$, `living' respectively in the ranges
of the orthogonal projectors $\hat{P}_g$ and $\hat{P}_e$. This is
a consequence of the fact that $[\hat{P}_m,\hat{C}_n]=0$,
$m=g,e\;$, $n=1,2,\ldots\ $. It is worth noting that the
Hamiltonian $\rotham_{12}$ can be regarded as the rotating frame
Hamiltonian of a trapped two-level ion in interaction with a laser
field characterized by the following parameters:
\begin{equation}\label{Effective_Parameters} \cases{
\omega_{12}\equiv&$\!\!
\omega_{13}-\omega_{32}=\omega_2-\omega_1$,\cr
\vec{k}_{12}\equiv&$\!\!\vec{k}_{13}-\vec{k}_{23}$.\cr }
\end{equation}
This effective coupling can be compared with the result found
performing the adiabatic elimination of the level $\Ket{3}$ (see
ref.~\cite{Steinbach}). We will come back to this point in the
next section.

\section{Dynamics of the Raman scheme}

The question of what the {\em complete} dynamics of the system is
now arises. First,
it will be convenient to adopt
the following notation. Given a couple of functions $f$ and $h$ of the
perurbative parameter $\lambda$, if
$f(\lambda)=h(\lambda) +O(\lambda^3)$,
we will write simply:
\[
f(\lambda)\app h(\lambda).
\]
Next, let us denote by
$\evo$ the evolution operator associated with the Raman scheme:
\begin{equation}
i\hbar\left(\frac{d}{dt}\,\evo\right)(t)=\hat{H}_\Lambda (t)
\,\evo(t),\ \ \ \evo(0)=\mathrm{Id}.
\end{equation}
Expressing $\evo$ in terms of the evolution operator associated with
the rotating frame Hamiltonian yields:
\begin{equation} \label{complete}
\evo (t)=
\hat{R}(t)\, \exp\!\left(-\frac{i}{\hbar}\rotham\, t\right).
\end{equation}
Now, according to what we have shown in the previous section, we have:
\begin{eqnarray}\label{Evolutor_1}
\nonumber
\hat{T}(t)
\!\!\! &:=& \!\!\! \exp\!\left(-\frac{i}{\hbar}\,\rotham\,t\right)\\
\nonumber
&=& \!\!\!
e^{-i\hat{Z}(\lambda)}\,
\exp\!\left(-i\Delta\,
e^{i\hat{Z}(\lambda)}\,\hat{H}(\lambda)\,
e^{-i\hat{Z}(\lambda)}\,t\right)
e^{i\hat{Z}(\lambda)}\\ \label{expr}
&\app&\!\!\!
e^{-i\left(\lambda\hat{Z}_1+\lambda^2\hat{Z}_2\right)}\,
e^{-i\Delta\left( \hat{H}_0+\lambda\hat{C}_1+\lambda^2\hat{C}_2 \right)t }\,
e^{i\left(\lambda\hat{Z}_1+\lambda^2\hat{Z}_2\right)},
\end{eqnarray}
where we have truncated the power expansions of
$\hat{Z}(\lambda)$ and $\hat{C}(\lambda)$ at the second order in $\lambda$.
Formula~{(\ref{expr})} provides an approximate expression of the
evolution operator in the remarkable form of a one-parameter group of
unitary transformations. Nevertheless, in order to achieve an
approximate expression allowing a direct comparison with the coarse grained
dynamics, we still need to perform some manipulation.
To this aim, observe that, since the commutators $[\lambda^2\hat{C}_2,
\lambda\hat{Z}_1+\lambda^2\hat{Z}_2]$ and
$[\lambda^2\hat{C}_2,\hat{H}_0+\lambda\hat{C}_1]=[\lambda^2\hat{C}_2,\lambda\hat{C}_1]$
are of the third order in $\lambda$,
maintaining our degree of approximation we can write
\begin{eqnarray}
\nonumber
e^{-i\left(\lambda\hat{Z}_1+\lambda^2\hat{Z}_2\right)}
e^{-i\Delta\left(\hat{H}_0+\lambda\hat{C}_1+\lambda^2\hat{C}_2\right)t}
\!\!\!&\app& \!\!\!
e^{-i\left(\lambda\hat{Z}_1+\lambda^2\hat{Z}_2\right)}
e^{-i\Delta\lambda^2\hat{C}_2 t}
e^{-i\Delta\left(\hat{H}_0+\lambda\hat{C}_1\right)t}
\\ \nonumber
&\app&\!\!\!
e^{-i\Delta\lambda^2\hat{C}_2 t}
e^{-i\left(\lambda\hat{Z}_1+\lambda^2\hat{Z}_2\right)}
e^{-i\Delta\left(\hat{H}_0+\lambda\hat{C}_1\right)t}.
\end{eqnarray}
Therefore, we can manipulate the second order expression~{(\ref{expr})}
of $\hat{T}(t)$ as follows:
\begin{eqnarray*}\label{Evolutor_2}
\hat{T}(t)
\!\!\! &\app & \!\!\!
e^{-i\Delta\lambda^2\hat{C}_2 t}
e^{-i\left(\lambda\hat{Z}_1+\lambda^2\hat{Z}_2\right)}
e^{-i\Delta\left(\hat{H}_0+\lambda\hat{C}_1\right)t}
e^{i\left(\lambda\hat{Z}_1+\lambda^2\hat{Z}_2\right)}
\\
&=& \!\!\!
e^{-i\Delta\lambda^2\hat{C}_2 t}
e^{-i\Delta\left(\hat{H}_0+\lambda\hat{C}_1\right)t}
\\
&\times& \!\!\!
e^{i\Delta\left(\hat{H}_0+\lambda\hat{C}_1\right)t}
e^{-i\left(\lambda\hat{Z}_1+\lambda^2\hat{Z}_2\right)}
e^{-i\Delta\left(\hat{H}_0+\lambda\hat{C}_1\right)t}
e^{i\left(\lambda\hat{Z}_1+\lambda^2\hat{Z}_2\right)}.
\end{eqnarray*}
Finally, we find a remarkable decomposition of $\hat{T}$:
\begin{equation}\label{Evolutor_3}
  \hat{T}(t)\app\hat{T}_{\mathrm{e}}(t)\,\hat{T}_{\mathrm{f}}(t),
\end{equation}
where we have set
\begin{equation}\label{Evolutor_Teff}
  \hat{T}_{\mathrm{e}}(t):=
  \exp\!\left(-i\Delta\left(\hat{H}_0+\lambda\hat{C}_1+
  \lambda^2\hat{C}_2\right) t\right),
\end{equation}
\begin{equation}\label{Evolutor_micro}
  \hat{T}_{\mathrm{f}}(t):=
  \exp\!\left(-i\left(\lambda\hat{Z}_1(t)+\lambda^2\hat{Z}_2(t)\right)\right)
  \exp\!\left(i\left(\lambda\hat{Z}_1+\lambda^2\hat{Z}_2\right)\right),
\end{equation}
with $\hat{Z}_k(t)\equiv
e^{i\Delta\left(\hat{H}_0+\lambda\hat{C}_1\right)t}\,\hat{Z}_k\,
e^{-i\Delta\left(\hat{H}_0+\lambda\hat{C}_1\right)t}$, $k=1,2$.
It is worth emphasizing that, by a completely analogous
procedure\footnote{The calculation may be carried on {\em
directly}, i.e. step by step as in the previous case but changing
the reordering of the exponentials in (\ref{Evolutor_1}) and
subsequent formulae, or from~(\ref{Evolutor_3}) exploiting the
fact that $\hat{T}(t)=\hat{T}(-t)^{\dagger}$.}, we get also
\begin{equation}\label{Evolutor_3_bis}
\hat{T}(t)\app\hat{T}_{\mathrm{f}}^\prime(t)\,\hat{T}_{\mathrm{e}}(t),
\ \ \ \hat{T}_{\mathrm{f}}^\prime(t):=\hat{T}_{\mathrm{f}}(-t)^\dagger.
\end{equation}
Notice that, due to the specific dependence of
$\hat{T}_{\mathrm{f}}$ on $t$, one has that
$\hat{T}_{\mathrm{f}}^\prime(t) \neq\hat{T}_{\mathrm{f}}(t)$.

In the light of formula~(\ref{Evolutor_3}) (or
(\ref{Evolutor_3_bis})), the time evolution in the rotating frame,
given by $\hat{T}$, may be thought of as a process consisting of
two fundamental components. One of these is an effective time
evolution, described by $\hat{T}_{\mathrm{e}}$, in which the
levels $\Ket{1},\Ket{2}$ are decoupled from the level $\Ket{3}$.
The other component, the one $\hat{T}_{\mathrm{f}}$ (or
$\hat{T}_{\mathrm{f}}^\prime$) is responsible for, is a correction
to $\hat{T}_{\mathrm{e}}$ and involves fast transitions (the
operators $\hat{Z}_1(t)$ and $\hat{Z}_2(t)$ oscillate at the
detuning frequency $\Delta$) from and to the third atomic level.
Considering the complete time evolution~{(\ref{complete})},
observe that the unitary evolution described by
$\hat{R}\,\hat{T}_{\mathrm{e}}$ corresponds to the effective
dynamics obtained in~\cite{Steinbach} restricting the analysis to
the coarse grained observables. In fact,
$\hat{R}\,\hat{T}_{\mathrm{e}}$ is the evolution operator
associated with the time-dependent effective Hamiltonian
\begin{equation}
\hat{H}_{\mathrm{e}}=
\hat{H}_{\mathrm{e}}^{(12)}+\hat{H}_{\mathrm{e}}^{(3)},
\end{equation}
where
\begin{eqnarray}
\hat{H}_{\mathrm{e}}^{(12)}(t)
\!\!\! &:=& \!\!\!
\hbar\nu\sum_{\alpha=x,y,z}(\hat{a}_\alpha^\dagger
\hat{a}_\alpha^{\phantom{\dagger}})
\otimes(\hat{\sigma}_{11}+\hat{\sigma}_{22})+
\hbar(\omega_1+\breve{\omega}_1)\hat{\sigma}_{11}
+\hbar(\omega_2+\breve{\omega}_2)\hat{\sigma}_{22}
\nonumber \\
\!\!\! & + & \!\!\! \left[\hbar
g_{12}e^{-i(\vec{k}_{12}\cdot\vec{r}-\omega_{12}t)}\hat{\sigma}_{12}
+ h.c.\right],
\end{eqnarray}
\begin{equation}
\hat{H}_{\mathrm{e}}^{(3)}\, :=\,
\hbar\nu\sum_{\alpha=x,y,z}(\hat{a}_\alpha^\dagger
\hat{a}_\alpha^{\phantom{\dagger}})\otimes\hat{\sigma}_{33}+
\hbar(\omega_3+\breve{\omega}_3)\hat{\sigma}_{33}.
\end{equation}
We remark that we have deduced this result analytically; no
adiabatic approximation has been performed. We also stress that a
relevant difference between $\hat{T}_{\mathrm{e}}$ and
$\hat{T}_{\mathrm{f}}$ consists in the kind of time dependence.
Indeed, on one hand, the unitary evolution $\hat{T}_{\mathrm{e}}$
forms a one-parameter group, hence it is expressible as the
exponential of a generator multiplied by $t$. It follows that a
truncated power expansion of the exponential retains its validity
only on a finite time span. On the the other hand,
$\hat{T}_{\mathrm{f}}$ can be expressed as the exponential of an
operator whose time dependence involves only sinusoidal factors.
In fact, we have:
\begin{eqnarray*}
\hat{T}_{\mathrm{f}}
\!\!\! &\app & \!\!\!
e^{-i\left(\lambda\left(\hat{Z}_1(t)-\hat{Z}_1\right)+
\lambda^2\left(\hat{Z}_2(t)-\hat{Z}_2\right)\right)}
\, e^{\frac{1}{2}[\lambda\hat{Z}_1(t)+\lambda^2\hat{Z}_2(t),
\lambda\hat{Z}_1+\lambda^2\hat{Z}_2]}
\\
&\app& \!\!\!
e^{-i\left(\lambda\left(\hat{Z}_1(t)-\hat{Z}_1\right)+
\lambda^2\left(\hat{Z}_2(t)-\hat{Z}_2\right)\right)}\,
e^{\frac{1}{2}\lambda^2[\hat{Z}_1(t),\hat{Z}_1]}
\\
& = & \!\!\!
e^{-i\left(\lambda\left(\hat{Z}_1(t)-\hat{Z}_1\right)+
\lambda^2\left(\hat{Z}_2(t)-\hat{Z}_2\right)\right)},
\end{eqnarray*}
where $[\hat{Z}_1(t),\hat{Z}_1]=0$ has been used.
It then follows that the truncated expansion
\begin{equation}\label{Linear_Tm}
  \hat{T}_{\mathrm{f}} \app
  1-i\lambda\left(\hat{Z}_1(t)-\hat{Z}_1\right)
 -i\lambda^2
  \left(\hat{Z}_2(t)-\hat{Z}_2\right)
 -\frac{1}{2}\lambda^2\left(\hat{Z}_1(t)-\hat{Z}_1\right)^2
\end{equation}
is legitimated independently on time.

Summarizing, we have shown that the standard adiabatic elimination
technique provides an effective dynamics, described by
$\hat{R}\,\hat{T}_{\mathrm{e}}$, that differs from the complete
second order dynamics of the Raman scheme for the presence of
another unitary evolution which can be cast in the form of the
exponential of a rapidly oscillating operator function of time.
Therefore, the factorization into a coarse grained and a fine
dynamics given by eq.~(\ref{Evolutor_3}) makes the correction to
the adiabatic approximation solution very readable and easy to be
calculated, in view of the expression~(\ref{Linear_Tm}). It is
worth noting that the corrections due to $\hat{T}_{\mathrm{f}}$
are small in amplitude, since the operators
$\lambda\hat{Z}_1,\lambda\hat{Z}_1(t)$ and
$\lambda^2\hat{Z}_2,\lambda^2\hat{Z}_2(t)$ are respectively of the
first and second order in the perturbative parameter. Moreover,
they provide terms oscillating at the detuning frequency. Hence,
as expected, the fine dynamics is small and fast. As a conclusive
remark, we wish to emphasize that this {\em micro-fast} behavior
brought to the light with the help of the method applied here
should become of practical interest in connection with time
resolution improvements of experiments.

\end{document}